\documentclass{biophys}
\usepackage{helvet,times}
\usepackage{bm,textcomp}

\jno{kxl014} 
\gridframe{N}
\cropmark{N}

\doi{doi: 10.1529/biophysj.106.090944}

\usepackage{amsmath}

\usepackage[round,numbers,sort&compress]{natbib}

\begin{document}

\setcounter{page}{1} 

\title{Threshold of microvascular occlusion: injury size defines the thrombosis scenario}

\author{Aleksey V. Belyaev$^{1,2}$,  Mikhail A. Panteleev $^{1,2,3,4}$ and Fazly I. Ataullakhanov $^{1,2,3,4}$}

\address{ 1). Center for Theoretical Problems of Physicochemical Pharmacology RAS, 38A
Leninsky Prospect, 119991 Moscow, Russia;
2). Federal Research and Clinical Center of Pediatric Hematology, Oncology and Immunology, 1 Samora Machel street,  117997 Moscow, Russia;
3). Department of Physics, M. V. Lomonosov Moscow State University, 119991 Moscow, Russia;
4). HemaCore LLC, 3 4th-8-Marta street, 125319 Moscow, Russia}



\begin{abstract}%
{Damage to the blood vessel triggers formation of a hemostatic plug, which is meant to prevent bleeding, yet the same phenomenon may result in a total blockade of a blood vessel by a thrombus, causing severe medical conditions. Here we show that the physical interplay between platelet adhesion and hemodynamics in a microchannel manifests in a critical threshold behavior of a growing thrombus. Depending on the size of injury, two distinct dynamic pathways of thrombosis were found: the formation of a non-occlusive plug, if injury length does not exceed the critical value, and the total occlusion of vessel by the thrombus otherwise.
We develop a mathematical model, which demonstrates that switching between these
regimes occurs as a result of a saddle-node bifurcation.
Our study reveals the mechanism of self-regulation of thrombosis in blood microvessels and explains
an experimentally observed distinctions between thrombi of different physical etiology.
This also can be useful for the design of platelet-aggregation-inspired engineering solutions.}
{Insert Received for publication Date and in final form Date.}
{Correspondence: $aleksey\_belyaev@yahoo.com$}
\end{abstract}

\maketitle

\section{INTRODUCTION}
Living systems at all levels of their organization display rich dynamic behavior, governed by various mechanisms of self-regulation, which are crucial for their functioning.
An interesting example of such phenomenon is hemostasis, which is aimed at prevention
of bleeding via aggregation of blood platelets and fibrin network formation \cite{Furie2005, Jackson}.
However, at certain circumstances, an overgrown intravascular aggregate, called `thrombus', may cause dangerous conditions, e.g. complete blockage of a blood vessel (vascular occlusion).
Experimental data \cite{Kurz1990, Denis2007, FurieNature, Stalker2013, Heemskerk2008, Oude1991} on thrombosis are highly
controversial, identifying complexity and hierarchy of involved physical and biochemical processes.
It is not yet understood, why some thrombi completely block the bloodstream leading to possibly catastrophic consequences \cite{Kurz1990, Denis2007}, and others accomplish their function not breaching the circulation \cite{FurieNature, Stalker2013, Heemskerk2008, Oude1991}.
Several suppositions emerged trying to explain these observations. The mechanism of self-regulation of thrombosis was ascribed to either biochemical reactions and platelet activation \cite{Heemskerk2008,FogelsonGuy2004}, or changing porosity of the thrombus \cite{Fogelson2013}, or its non-uniform structure \cite{Stalker2013, Jackson2, Volpert2013}, but is still being debated. Early studies demonstrate that thrombosis is governed mainly by two competing factors:
the rate of platelet attachment from bloodstream and the intensity of hydrodynamic
forces that prevent platelets from adhering to the
thrombus \cite{BegentBorn, Richardson1973, Sato1986, Oude1991, Pivkin2006, AlberSoftmatter2009, Volpert2013}.
It was revealed that platelet aggregation rate does not simply increase with blood flow velocity,
yet exhibits a maximum with a subsequent decrease due to growing hydrodynamic forces that inhibit platelet adhesion \cite{BegentBorn, Richardson1973, Tokarev2011a}. The combination of hydrodynamic features of microvasculature with nonlinear shear-dependent platelet aggregation rate, in principle, may stop the growth of thrombus. In this study we use
mathematical modeling to check validity of this hypothesis and focus on principal physical effects that drive thrombosis.

\begin{figure}
  \centerline{\includegraphics[ ]{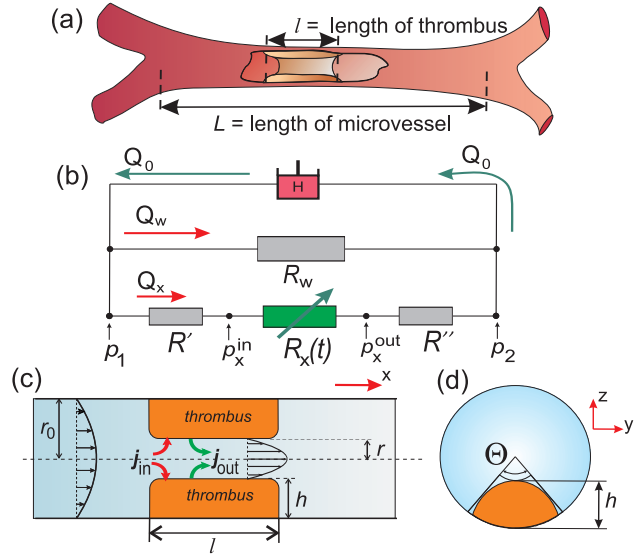}}
  \caption{(Color online) Scheme of the model. (a) Sketch of blood vessel with thrombus. (b) Hydraulic scheme of circulatory system. Thrombotic vessel is depicted as series of resistances $R'$, $R_x$ and $R''$; blood can bypass through remaining heathy microvasculature $R_w$. (c) Schematic side view of a thrombotic vessel with axisymmetric thrombus: $r(t)$ is the radial coordinate of the thrombus's apex, $h$ is the height of the thrombus, $r_0$ is the radius of the vessel. (d) Schematic cross-section of a thrombotic vessel: $\Theta$ is the angular size of thrombus. Axisymmetric thrombus in panel (c) is a particular case, corresponding to $\Theta=2\pi$.   }
  \label{fig1}
\end{figure}

\section{METHODS}
\subsection{Blood flow}
We account for closure of blood circulatory system and deduce hydrodynamic conditions within the thrombotic vessel (Fig.1(a)). Heart acts as a pump, providing a systole-averaged constant flow rate $Q_0$ (Fig.1(b)). We neglect pulsations in microvasculature because of their small relative amplitude (\cite{Fung_book, Caro, Oude1991}).
Thrombotic vessel consists of a stenosed segment (with resistance $R_x$) and two healthy ones ($R'$ and $R''$). A large number of healthy vessels with overall resistance $R_w\sim 100$ Pa$\cdot$s/cm$^3$ \cite{Heart} provide a shunt through which blood can bypass thrombotic vessel.
In our study, $R_w$ is assumed constant in time, as we neglect the flexibility of vessel walls in microvasculature with respect to stenosis caused by the thrombus, so radii of healthy vessels remain unchanged.
The resistance of bigger vessels may be neglected compared to that of microvasculature.  We restrict analysis to arterioles and venules with diameters smaller than 1000 microns and bigger than 50 microns, for which Reynolds numbers are small ($Re < 0.1$) \cite{Caro}, so flows are governed by quasi-stationary Stokes equations. For such vessels we also have no option of entanglement of red blood cells in the fibrin network, which is typical for venous thrombi.

We approximate blood rheology in our model as Newtonian, i.e. characterized by a Navier-Stokes model. This should be treated as a simplifying assumption for initial study. However, we expect it to be a reasonable approximation for high shear rates ($> 100$ 1/s), provided by the data from Ref. \cite{Fung_book_2}.

Let $L$ be the length of a thrombotic blood vessel measured between two consequent bifurcations. Healthy segments have circular cross-section, and the lumen shape of the stenosed one is disturbed by thrombus, Fig.1(c,d). In our numerical model the actual shape of thrombus is approximated by cylindrical segment, protruded into the vessel and characterized by $l$, $r$ and $\Theta$. The angular size of the thrombus $\Theta$ and its height $h$ are allowed to change in time, while $l=const$. Resistance of damaged segment $R_x(t)$ changes due to growing stenosis, and consequent changes happen to pressure and flow rate in whole circulatory system.
Pressure difference $\Delta p_{12}=p_1-p_2$ between vessel ends may change due to thrombus, as well as pressure drop across thrombus $\Delta p_x= p_x^{\rm in} -p_x^{\rm out}$ and flow rate $Q_x$ through the vessel \cite{FogelsonGuy2004}. Thus, hydrodynamics and thrombus growth are coupled and should be considered within a unified self-consistent model. Within our approach at each timestep for a given thrombus's configuration hydrodynamic values (shear rate, $\Delta p_x$ and $Q_x$) were found numerically as a solution of linearized Navier-Stokes equation for incompressible flow (for details see sections 1-5 in \cite{SIText}), then thrombus shape was adjusted according to hydrodynamics-dependent growth rate.

\subsection{Platelet accumulation}
Platelets circulate in the blood relatively inactive and
do not adhere to healthy vessel walls. Damage of vessel wall starts a cascade of processes leading to formation of a thrombus  \cite{Furie2005}. Its volume increases due to  attachment of platelets supplied by the bloodstream. Platelet adhesion at early stage is reversible \cite{Jackson, Stalker2013}, and thrombus growth rate is the balance of rates of platelet attachment $j_{\rm in}$ and detachment $j_{\rm out}$:
 \begin{equation}\label{dV}
    \frac{d }{dt}V_{\rm thromb}= \int_{\rm surf} \left( j_{\rm in} - j_{\rm out} \right) dA,
\end{equation}
where the integral is taken over the surface of thrombus capable to catch platelets. Platelet attachment indeed depends on hydrodynamic conditions \cite{BegentBorn, Richardson1973, Turittoetal, Baumgartner, SatoOhshima,  Zhao2011, Tokarev2011a}. In our model, moving RBCs influence platelet transport towards vessel walls based on the concept of shear-enhanced diffusivity \cite{TurittoBaumgartner, TurittoBaumgartner2, TurittoBaumgartner1979,  Tokarev2011a, BarkKu}, implicitly expressed as a shear-dependent platelet accumulation rate $j_{\rm in}$. We use a power-law function proposed earlier $j_{\rm in} \simeq \psi \dot{\gamma}^{1-\beta}$ with $\beta \approx 0.2$ for human blood cells.
This expression comes from the supplementary materials in \cite{Tokarev2011a} and also from rigorous mathematical analysis by \cite{Diamond1997}. Platelet flux towards the wall is determined by the frequency of their near-wall inelastic collisions with erythrocytes (RBCs), as proposed in \cite{Tokarev2011a} and verified therein by comparison with experiments \cite{TurittoBaumgartner, TurittoBaumgartner2, TurittoBaumgartner1979}. While collisions of platelets with platelets may be neglected (only collisions of platelets with RBCs taken into account), $j_{\rm in}$ reads:
\begin{equation}
  j_{\rm in} \approx \varepsilon(\dot\gamma, \: d_{\rm pl}/d_{\rm RBC} ) \cdot K d_{\rm RBC} \Phi \dot\gamma c_{\rm pl},
\end{equation}
where   $c_{\rm pl}$  is platelet concentration in blood near the vessel wall, $K$ is the coefficient dependent on shape and size of platelets and RBCs, $d_{\rm RBC}$ and $d_{\rm pl}$ are diameters of platelets and RBCs respectively, $\Phi$ is the hematocrit, and $\dot\gamma$ is the shear rate at the vessel wall. The collision efficiency $\varepsilon(\dot\gamma, \: d_{\rm pl}/d_{\rm RBC} )$ , as a function of shear rate, has the power-law form (see Eq.(S14) in supplementary data to Ref.\cite{Tokarev2011a}):
\begin{equation}
  \varepsilon(\dot\gamma, \: d_{\rm pl}/d_{\rm RBC} ) =  A \cdot \left(\frac{2.725}{\dot\gamma}\right)^\beta,
\end{equation}
where coefficient $A$ and index $\beta$ both depend only on cell diameters' ratio $d_{\rm pl}/d_{\rm RBC}$.
Thus, recasting this expressions, we get the formula $j_{\rm in} \simeq \psi \dot\gamma^{1-\beta}$ that we use in the present paper.
Here  $\psi$ is a parameter defined by the form and size of platelets and RBCs, near-wall concentration of platelets in blood and hematocrit. Index $\beta$ depends on the sizes and mechanical properties of RBCs and platelets \cite{Tokarev2011a}.

Thrombus erosion rate $j_{\rm out}$ should be proportional to viscous shear stress per unit area of thrombus $\eta \dot{\gamma}$, where $\eta$ is the viscosity of blood plasma, inversely proportional to inter-platelet adhesive forces and should depend on shape and density of platelet aggregate:  $j_{\rm out}= \xi_d \dot{\gamma}$,
where $\xi_d$ is a constant with a dimensionality of length, which accounts for all non-hydrodynamic effects.
The effective growth rate $k_{\rm eff}= \left( j_{\rm in} - j_{\rm out} \right)$ is a power function of wall shear rate:
\begin{equation}\label{Keff_beta}
    k_{\rm eff}= \psi \dot{\gamma}^{1-\beta} - \xi_d \dot{\gamma}.
\end{equation}
This value is positive for small $\dot{\gamma}$, but changes its sign if $\dot{\gamma}>\dot{\gamma}_{\rm cr}$, where $\dot{\gamma}_{\rm cr}= (\psi/\xi_d)^{1/\beta}$ corresponds to balance between the influx and the outflux of platelets ($k_{\rm eff}=0$).
Expression (\ref{Keff_beta}) is an approximate formula, but it reflects the experimentally observed behavior of platelet accumulation with respect to shear rate change. The adhesion is absent without flow and increases initially with the increasing shear rate. However, for further increase of shear rate, the flow starts rupturing the thrombus \cite{Stalker2013,  Furie2005, FurieNature}, which leads to lowering of platelet accumulation rate for high ($>10^4$ 1/s) shear rate \cite{Turittoetal, TurittoBaumgartner1979, BegentBorn}.

In particular, for axisymmetric thrombus ($\Theta=2\pi$) the model permits analytical consideration.
When thrombus grows over the whole perimeter of vessel's cross-section, the lumen has a circular shape with inner radius $r(t)<r_0$.
Since the only variable that determines the size  of an axisymmetric thrombus is the inner lumen radius $r(t)$, in that case Eq.(\ref{dV}) reads
\begin{equation}\label{dV_sym}
   - 2\pi r l \cdot \frac{d r}{dt}= k_{\rm eff}(\dot{\gamma}) \cdot A,
\end{equation}
where $A\simeq 2\pi r l$ is the area capable of accumulating platelets, and $\dot{\gamma}=  \Delta p_x(t) \cdot r(t)/(2\eta l)$ is the shear rate at thrombus's top surface (see section 4 of \cite{SIText}).

For convenience we introduce the following dimensionless values that determine thrombosis dynamics:
the relative thrombus length $\lambda= l/L$, scaled vessel's radius $\rho_0= r_0/ L$, inner radius of the lumen $\rho= r/ L$, the adhesion parameter $\xi=(\dot{\gamma}_0/\dot{\gamma}_{\rm cr})^{\beta}=\xi_d \cdot\dot{\gamma}_0^\beta / \psi$ that characterizes stability of platelet aggregates subjected to hydrodynamic forces in a vessel, and $\Psi=\psi \dot{\gamma}_0^{1-\beta}\tau/L$ that determines the typical timescale $\tau$ for dynamics of thrombosis. We define typical scale for hydraulic resistance as:
\begin{equation}
  [R]= \frac{8 \eta}{\pi L^3}.
\end{equation}
As for the flow rate scale, it is convenient to express this value through combination of blood flow rate at the outlet of the heart $Q_0$ and the shunting hydraulic resistance $R_w$: $[Q]=Q_0 R_w/[R]$.
Pressure difference is thus scaled to a factor $[\Delta p]=Q_0 R_w$.
Shear rate in the stenosed vessel $\dot{\gamma}$ was scaled to the initial shear rate (when no thrombus yet grown)
\begin{equation}
  \dot{\gamma}_0= \frac{[\Delta p]\rho_0}{2\eta}= \frac{Q_0 R_w r_0}{2\eta L},
\end{equation}
so that dimensionless shear rate $G=\dot{\gamma}/\dot{\gamma}_0$.

\begin{figure}
 \centerline{\includegraphics[ ]{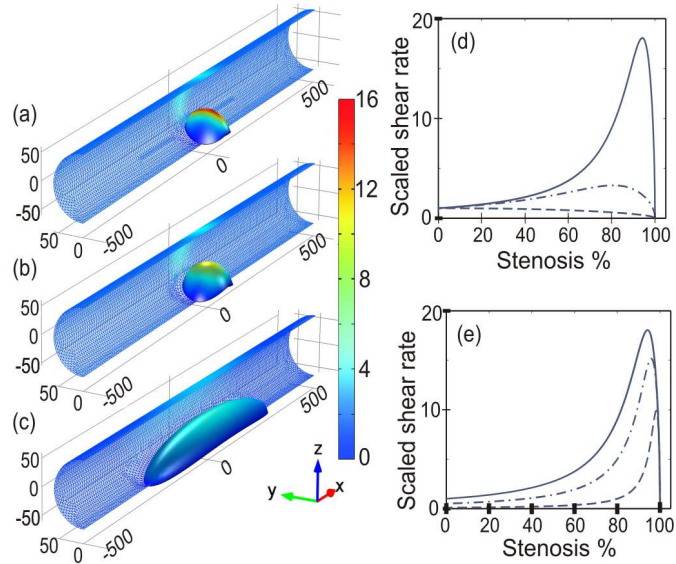}}
  \caption{(Color online) (a-c) Mapping of scaled shear rate $\dot{\gamma}/\dot{\gamma}_0$ over the thrombus's surface and walls of the blood vessel computed with finite-difference simulation for different lengths of a thrombus: 50 $\mu$m (a), 100 $\mu$m (b) and 500 $\mu$m (c). Vessel diameter is 100 $\mu$m. Pressure drop between vessel's ends $\Delta p_{12}=100$ Pa.
  (d)   Dependence of shear rate on stenosis degree for $l/L=0.01$ (solid),  $0.1$ (dash-dot) and $0.75$ (dash). Here $R_w/[R]=10^{-3}$, $[R]=8 \eta/(\pi L^3)$.
  (e)  Shear rate vs. stenosis for
  different hydraulic resistances of the shunt: $R_w/[R]=10^{-3}$ (solid), $1$ (dash-dot), $10$ (dash);
  thrombus length $l/L=0.01$.}
  \label{fig2}
\end{figure}

We note here the simplifying assumption that thrombus growth is solely due to platelet aggregation in our model. Our model does not consider explicitly the biochemistry of blood coagulation that is indeed important \cite{anand2006} for clot formation. However, our approximation assumes that these effects are taken into account implicitly in the rate of platelet attachment estimated from the experimental data. What is important is the balance between the mechanical properties of the aggregate and hydrodynamic forces acting to disrupt it. Besides that, recent reports \cite{Stalker2013} suggest that general shape of thrombi and sub-occlusive thrombus stabilization is completely determined by the dynamics of thrombus "shell" containing no fibrin/thrombin and composed of platelets with only slight activation.

\section{RESULTS}

We performed numerical simulations in order to reveal the mechanism of thrombosis regulation.
Fig.2 suggests that longer thrombus experiences lower shear stress on its surface.
In Fig.2(a-c) the maximal shear rate (and, thus, the force)
on top of the plug gradually reduced 4-fold, when length of thrombus  was increased from 50 $\mu$m to 500 $\mu$m.
Our analytical calculations for axisymmetric case ($\Theta_0=2\pi$) also support these findings, as illustrated in Fig.2(d). Note the nonmonotonic dependence of shear rate on the degree of stenosis: when $\lambda \ll 1$, shear rate increases with growing stenosis, reaches maximum $\dot{\gamma}_{\rm max}$ and then goes to zero  as occlusion occurs. That is consistent with earlier observations \cite{Sato1986, Diamond2012_2}.  Hydrodynamic shear forces $f_h \propto \dot{\gamma}$ may prevent occlusion, if $\dot{\gamma}_{\rm max}$ is high enough. Therefore, longer thrombi tend to be more occlusive. Increase of relative shunting resistance lessens $\dot{\gamma}_{\rm max}$, Fig.2(e), yet not changing qualitative picture of thrombosis.
For realistic system the limit $R_w \rightarrow 0$ (equivalent to $\Delta p_{12}=const$) is a good approximation (see section 5 in \cite{SIText}). We would like to notice the fact that Fig.2(e) shows the ratio of actual dimensional shear rate at walls of a stenosed region to the initial wall shear rate (when thrombus's height is zero). For that reason, this plot resembles the \textit{relative} changes due to stenosis respective to the initial flow. While the smaller absolute values of shear rate in stenosed vessel are expected for smaller $R_w$,  the relative change of shear rate due to growing stenosis is much better pronounced in that case. In other words, smaller $R_w$ stands for higher sensitivity of the system to growing stenosis.

A good consistency with experimental data attests validity of our theory, Fig.3(a).
We found model parameters $\psi$, $\xi$ and $L$  by fitting experimental data from \cite{Sato1986} (for details see section 8 and Fig.S5 in \cite{SIText}).
Specifically, we found that $\dot{\gamma}_{\rm cr}  \approx 3400$ $s^{-1}$ corresponds to the balance between the influx and the outflux of platelets in this particular experiment.

Thrombus growth rate $k_{\rm eff}$ can be positive, or negative, depending on system's parameters, so that thrombus can either grow or decrease in size. If we plot this value as a function of the thrombus's height, Fig.3(b), this effective growth rate diagram would characterize dynamics of thrombosis. We found that a saddle-node bifurcation takes place, when relative thrombus length $\lambda$ is changing. If $\lambda$ is above critical the only stable point exists, corresponding to total occlusion. As $\lambda$ decreases, a saddle-node fixed point emerges and then divides into two equilibria points. The one with smaller $h/r_0$ is a locally stable node corresponding to a non-occluding thrombus, and another one is not stable.

Typical time courses for axisymmetric thrombi are presented in Fig.3(c) for different $\lambda$.
The analysis shows that there exists a critical thrombus length $\lambda_ {cr}$, which differentiates the regimes of occlusion and thrombosis stoppage. For $\lambda> \lambda_ {cr}$ the occlusion occurs in a finite period of time, and for $\lambda< \lambda_{\rm cr}$  thrombus reaches finite size.
The value of $\lambda_{\rm cr}$ should be determined from
the condition $\dot{\gamma}_{\rm max} = \dot{\gamma}_{\rm cr}$,
where $\dot{\gamma}_{\rm max}$ is the peak value of dimensionless shear rate
and $\dot{\gamma}_{\rm cr}$ is the shear rate for which average influx and outflux of platelets in Eq.(\ref{dV}) are balanced and $k_{\rm eff}=0$. This requirement leads to the equation (see \cite{SIText} section 9 for mathematical details):
\begin{equation}\label{separatrix}
     \lambda^3 (1-\lambda) = \frac{27}{256} \left( \frac{\dot{\gamma}_0}{\dot{\gamma}_{\rm cr}} \right)^{4}.
\end{equation}
Note that $\dot{\gamma}_{\rm cr}$ is the biological characteristic, that depends on adhesive properties of platelets, hematocrit, sizes of cells, while $\dot{\gamma}_0$, the initial wall shear rate, is the hydrodynamical characteristic of the system, that is only dependent on geometry and pressure.
Solution of Eq.(\ref{separatrix}) determines critical relative thrombus length $\lambda_{\rm cr}=l_{\rm cr}/L$, so that
if $\lambda> \lambda_{\rm cr}$, the thrombosis always proceeds to occlusion,
and for  $\lambda< \lambda_{\rm cr}$ the thrombus reaches stable non-occlusive size. A saddle-node bifurcation occurs at $\lambda = \lambda_{\rm cr}$.

Fig.3(d) summarizes our findings for axisymmetric thrombus in a form of a parametric diagram. $X$-axis of the diagram shows $\lambda=l/L$, and $y$-axis corresponds to $\dot{\gamma}_{\rm cr}/\dot{\gamma}_0$.
The upper-right corner of the diagram corresponds to parameters,
for which the thrombus grows to occlusion.
Domain of thrombosis stoppage is enclosed between line  $\lambda_{\rm cr}=f(\dot{\gamma}_{\rm cr}/\dot{\gamma}_0)$, found from Eq.(\ref{separatrix}), and the horizontal line $\dot{\gamma}_{\rm cr}=\dot{\gamma}_0$.
Below the former one any platelet aggregate
will be washed away by the bloodstream, as $\dot{\gamma}_0 > \dot{\gamma}_{\rm cr}$.

\begin{figure}
  \centerline{\includegraphics[ ]{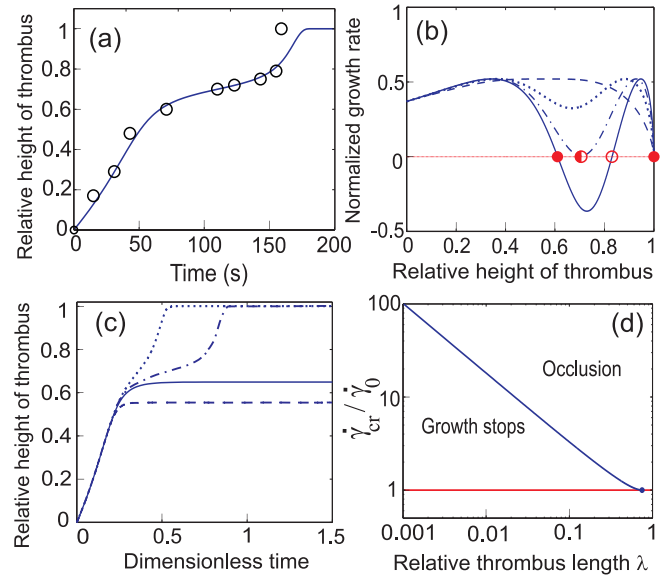}}
  \caption{(Color online) (a) Time course for changes in relative thrombus height in venule with inner radius $r_0 =31$ $\mu$m, length of injury $l= 4 r_0$; $\dot\gamma_0 \approx 320$ $s^{-1}$. Theory (solid line) is compared to experiments from \cite{Sato1986} (circles); error bars are smaller than symbols. Dimensionless model parameters found from the fitting: $\Psi= 0.026$, $\xi= 0.63$, $r_0/L = 5.8 \cdot 10^{-3}$. We also set $\beta=0.2$ for all theoretical curves. (b) Phase portrait for the growth of axisymmetric thrombus: the normalized effective thrombus growth rate $k_{\rm eff}/ (\dot\gamma_0^{1-\beta}\psi)$ as a function of the ratio of thrombus's height to vessel's radius $h/r_0$ for $\xi=0.63$ and $r_0/L= 0.05$. Solid, dash-dot, dotted and dashed lines correspond respectively to $\lambda=0.016$, $0.022$, $0.035$ and $0.100$. Filled and empty circles mark stable and unstable equilibrium points respectively, and the semi-filled circle is the saddle-node. (c) Growth dynamics of axisymmetric thrombi: for $\lambda= 0.028$ (dotted line) and $0.023$ (dash-dot) occlusion occurs, while for $0.020$ (solid line) and $0.005$ (dash) the thrombus reaches a quiescent height and stops growing. Here $\Psi= 0.05$, $\xi= 0.63$, $r_0/L = 0.01$. Critical size of injury we found $\lambda_{\rm cr} \simeq 0.022$. (d) Diagram of axisymmetric thrombosis: normalized critical shear rate $\dot{\gamma}_{\rm cr}/\dot{\gamma}_0$  vs. relative thrombus length $\lambda$. }
  \label{fig3}
\end{figure}

Fig.4(a) suggest that non-symmetric thrombi show more tendency to occlusion than the ring-shaped ones.
We found that thrombi with initial angular size $\Theta_0$ close to $2\pi$  stopped their growth, since $\lambda< \lambda_{\rm cr}$, yet injuries with smaller $\Theta_0$ and same $\lambda$ led to occlusion.
The threshold still remains for non-symmetric case, as shown in  Fig.4(b), but $\lambda_{\rm cr}$ becomes smaller for smaller $\Theta_0$.
According to our results, the highest possible shear rate is attained for circular cross-section (for mathematical details see section 3 in \cite{SIText}), thus ring-shaped thrombus leads to the most adverse conditions for occlusion.

\begin{figure}
  \centerline{\includegraphics[ ]{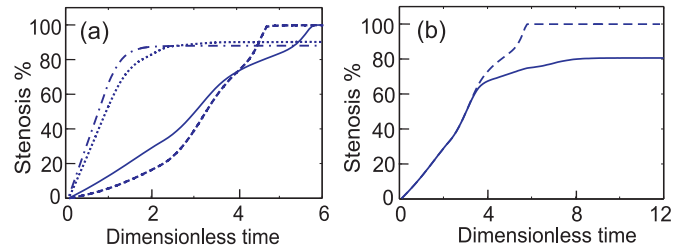}}
  \caption{(Color online) Dynamics of non-symmetric thrombi
  (a) Time courses for thrombi with initial angular sizes
$\Theta_0=\pi/3$ (solid), $\pi/10$ (dashed), $3\pi/2$ (dash-dot)
and $2\pi$ (dotted). Thrombus
length $\lambda=0.018$ for all curves, this value is smaller than $\lambda_{\rm cr}$ for
axisymmetric case.
(b) Time courses for thrombi
with $\Theta_0=\pi/3$, $\lambda=0.018$ (dash) and $0.005$ (solid).
$\Psi= 1.0$ and $\xi= 0.63$.}
\label{fig4}
\end{figure}

The only dimensionless parameter defining details of platelet adhesion, is $\xi=( \dot{\gamma}_0 /  \dot{\gamma}_{\rm cr})^\beta$. It quantifies the ratio of platelet
detachment to accumulation rates. Fig.5(a) shows that higher $\xi$ makes occlusion less plausible. Final size of the thrombus is determined by $\xi$, see Fig.5(b). Oppression of platelet adhesion (increase of $\xi$), diminishes the size of a stable thrombus when $\lambda<\lambda_{\rm cr}$.  Notice that `occlusive' equilibrium is stable for any $\lambda$.

\begin{figure}
  \centerline{\includegraphics[ ]{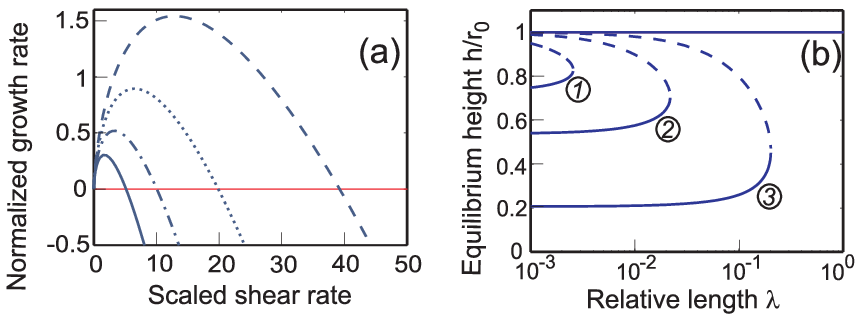}}
  \caption{(Color online) Effect of the adhesion strength parameter $\xi$ on thrombosis.
  (a) Normalized effective growth rate $k_{\rm eff}/ (\dot\gamma_0^{1-\beta}\psi)$ versus $\dot{\gamma}/\dot{\gamma}_0$:  $\xi=0.72$ (solid), $0.63$ (dash-dot), $0.55$ (dotted), and $0.48$ (dash).
  (b) Bifurcation diagram for axisymmetric thrombus: solid and dash lines show locally stable and unstable thrombus heights respectively in dependence on $\lambda$; $\xi \simeq 0.46$, $ 0.63$ and $0.87$ for curves \textit{1}, \textit{2} and \textit{3}. }
  \label{fig5}
\end{figure}

\section{DISCUSSION}

Our results suggest that the mechanism of thrombus growth regulation in microvasculature rests on the interplay between the non-trivial hydrodynamics-dependent platelet aggregation law Eq.(\ref{Keff_beta}) and features of microcirculatory network.
Namely, initial prevalence of adhesion favors the formation of a hemostatic plug, yet consequent increase of shear forces impede thrombus growth and may eventually stop it. However, if injury length exceeds the critical value, this mechanism fails to prevent occlusion due to lesser shear rates.  This explains why ferric chloride-induced thrombosis characterized by vast injuries is always occlusive \cite{Kurz1990, Denis2007}, while small laser-induced thrombi may stop growing \cite{Stalker2013}. Interestingly, recanalization of occlusive thrombi \cite{Reidel_et_al, Strbian_et_al} and the risks of implant-induced thrombosis \cite{stent2010} also depend on the thrombus length in the same threshold manner.

The principal conclusion is an existence of a threshold injury length $l_{\rm cr}=\lambda_{\rm cr}L$ demarcating regimes of thrombosis due to occurring saddle-node bifurcation. This value depends on relative  adhesion strength expressed via fraction of $\dot{\gamma}_{\rm cr}= (\psi/\xi_d)^{1/\beta}$  and initial wall shear rate $\dot{\gamma}_0$, Eq.(\ref{separatrix}).
Another interesting thing is a noticeable bistability, Fig.5(b), predicted for subcritical $\lambda$. The reasons for the observed dynamics are contained within the system itself. The positive feedback is provided by shear-dependent platelet margination, manifested in the aggregation rate $j_{\rm in}$ increasing with shear rate. The shear rate peak effect, Fig.2(d), serves as a mechanism to filter out small stimuli, and acts together with thrombus erosion $j_{\rm out} \sim \dot{\gamma}$ to inhibit thrombosis (i.e. negative feedback) and prevent `explosive' growth. We note that such behavior comes naturally from the hydraulics of the system and is independent from the exact nature of platelet transport. What is essential is that $j_{\rm in}$ grows slower with shear rate than $j_{\rm out}$ does. Even without any collision-induced platelet margination, classical transport theories \cite{ZydneyColton, TurittoBaumgartner, TurittoBaumgartner2, TurittoBaumgartner1979} imply that platelet
transport to the wall will increase with shear rate as $\dot\gamma^{\alpha}$ with $\alpha<1$, so collision-induced margination is not principal for our results.

Remarkably, our conclusions complement the results of \cite{Beltrami_Jesty2001}, who predicted mathematically that there exists a threshold size of a membrane patch for proteolysis and enzyme (thrombin) production during blood plasma coagulation. This fact discloses the fundamental generality of the described flow-dependent regulation mechanism. Apparently, in both systems the thresholds originate from the presence of feedbacks.

For a longer thrombus, depletion of platelets due to their
attachment at the upstream region of the thrombus may become
important. Thus, axial variations in its
height are likely, which in turn affects the shear rate distribution
on the thrombus and, potentially, its further growth. However, if so, the hydrodynamic (shear) forces at thrombus's surface would be greater in the upstream region with higher stenosis, decreasing local growth rate $k_{\rm eff}$, and so leveling the height of a plug. \textit{In vivo} experiments (see supplementary in \cite{Stalker2013}) support this idea: loosely packed upper layers of thrombus are facilely smeared by the flow. In any case, the occlusion event would be governed by the region of highest stenosis, for which our model should give qualitatively correct results.

Our analysis is restricted to platelet thrombi in venules
and arterioles and does not consider larger blood vessels. In
the latter case one should correct the hydrodynamic values to account for non-laminar flows \cite{Fung_book}. Furthermore, as we focus on the effects caused by hydrodynamics, the difference between arterioles and venules from this point
is rather quantitative \cite{Fung_book}, and no specific modification is required to our model in order to switch between these types of vessels.
The presented theory does not describe the stage of initiation of thrombus growth and the lag-time, which is not the aim of presented research.
We also do not consider explicitly a number of biological processes and features (platelet activation, fibrin growth, granule secretion, extent of damage and tissue factor concentration, etc.) that take place inside the "core" \cite{Stalker2013} of the thrombus
and basically can not alter the considered surface-related hydrodynamic mechanism.
These processes naturally addressed to strengthen the plug \cite{Jackson}  would correct quantitative estimations of thrombus growth dynamics, however, would not principally change hydrodynamic effects, which are central in our study.

Our model is inspired by experimental facts about thrombus's structure, in particular by Ref.\cite{Stalker2013}, which clearly shows that platelets attach to and detach from relatively thin interfacial layer, named "shell", in contrast to "core", which consists of firmly attached platelets, partly covered with fibrin. Within the "shell", which has a thickness of several platelets, the whole parts of clot (whole platelet aggregates) can detach, not only the individual cells. Note, that our formulation includes this fact, since the detachment rate is described in our model by a parameter $\xi_d$, which possibly can be varied in a wide range.

Finally, length of the plug may change during the thrombosis \cite{Stalker2013,Furie2005}.
However, this elongation is usually moderate (smaller than order of magnitude) and is not radical for microvascular thrombi, provided by experimental evidences \cite{Diamond2011} that the adhesive substrate/activator area limits the length of the resulting thrombus for small concentrations of tissue factor at the injury site. It was shown experimentally, that an assumption of constant thrombus length is quite adequate as long as activation is not very strong (tissue factor concentration is moderate, compare to results presented in \cite{OkorieDiamond2008}). Therefore, our analysis is precise if injury is not very severe, and concentration of activators near the injury site is moderate. We believe that for a strong activation the exact value of critical length $\lambda_{\rm cr}$ could change, nevertheless our qualitative results would not be principally affected. Besides that, since the dependence of vascular resistance $R_x$ of the thrombotic region is dependent on length $l$ much weaker than on lumen radius $r$ (powers of 1 and 4 respectively, see section 4 of Supplementary file), one may expect that main effect on the blood flow (and thus on platelet accumulation rate) would be from the changes of lumen radius $r$, while alterations of $l$ are secondary in the considered system. This can be seen in Fig.2(d): to switch from solid line to dash-dotted line we have to change $\lambda=l/L$ through the whole order of magnitude (from 0.01 for solid to 0.1 for dash-dotted). However, qualitatively, the curve for shear rate has the same hillock shape. And for $\lambda<0.01$ these curves change even smaller: we cannot distinguish between lines for 0.005 and 0.01 in this figure. Since shear rate increase is the principal regulator for the occlusion event, our assumption of constant thrombus length is expected to give reasonably precise results, while the injury is sufficiently small ($\lambda \leq 0.01$).

Several effects predicted by our theory are not obvious. The most direct approach to experimental verification is to study thrombus formation under constant pressure conditions using different sizes of the damaged area. We stress that shunt vessels $R_w$ should be necessary implemented in experimental setup to correctly reproduce microvascular hemodynamics and thrombosis \textit{in vitro}. Our model predicts that there should be a triggering from occlusion to stable thrombus for a certain length of injury. Another way is to lower $\dot{\gamma}_{\rm cr}/\dot{\gamma}_0$ by using platelets with partial adhesion receptor deficiencies, or increasing pressure $\Delta p_{12}$.

\section{CONCLUSION}

In present work with usage of only basic physical principles, not involving complex biochemical concepts, we quantitatively described the switching mechanism in microvascular thrombosis. Our findings resolve fundamental contradictions in experiments and constitute an essential step towards understanding and treatment of different thrombotic disorders. Our results are widely applicable outside the hemostasis field. For example, blood-clotting-inspired colloidal-polymer composite systems \cite{AlexanderKatz_nature2013} are an attractive class of new smart materials with tunable properties. The dynamics of growth of these artificial plugs depends mostly on the shear rate and the strength of the polymer-colloid binding potential, and no biochemical activation is required. We believe, that our theory describes such artificial clotting systems as well, providing a basis for their usage in numerous applications, e.g. microfluidic devices.

\section{AUTHOR CONTRIBUTIONS}
A.V.B. developed a model, carried out the calculations, analyzed results;
M.A.P. analyzed the results; F.I.A. outlined the scientific problem and analyzed the results.
All authors reviewed the manuscript.

\section{ACKNOWLEDGEMENTS}
The study was supported by Russian Science Foundation (grant number 14-14-00195) with the exceptions of CFD simulations of blood flow in a thrombotic vessel and calculations of growth dynamics for non-axisymmetric thrombus, which were supported by the Stipend of President of Russian Federation for young scientists (grant number SP-2427.2015.4).

\section*{SUPPLEMENTARY MATERIAL}

An online supplement to this article can be found by visiting BJ Online at http://www.biophysj.org.

\bibliography{thrombus}

\bibliographystyle{biophysj}


\newpage




\end{document}